\documentclass[12pt]{iopart}

\usepackage{graphicx} 
\begin{document}

\title{The low energy frontier: What is exciting about physics below the top RHIC energy}

\author{Marcus Bleicher}

\address{Institut f\"ur Theoretische Physik, Goethe Universit\"at
Frankfurt,Germany} \address{Frankfurt Institute for Advanced Studies
(FIAS), Ruth-Moufang-Str. 1, 60438 Frankfurt, Germany}

\ead{bleicher@th.physik.uni-frankfurt.de}

\begin{abstract} These proceedings summarize my plenary talk at Quark Matter 2011 with a focus on the future perspectives of the low energy programs at RHIC, FAIR, NICA and CERN.  \end{abstract} \maketitle

\section{Introduction}

Over the last decade relativistic heavy ion physics has made tremendous advances in terms of our understanding of the phase diagram of Quantum-Chromo-Dynamics (QCD). Most of these advances are, however, related to the high temperature and small baryon density regime as encountered in collisions at the top RHIC energy. Here one has been able explore the properties of the QCD-matter created in great detail and to estimate its transport properties quite accurately from comparisons between viscous hydrodynamics calculations and experimental data (especially on the elliptic flow, $v_2$ and on the attenuation of particles with high transverse momenta, called jet quenching).  This energy regime offers certain advantages and disadvantages, e.g. perturbative QCD methods may work reliably for the first time in heavy ion reactions, and straightforward hydrodynamics seems to allow for a good description of the experimental data. While on the experimental side, the luminosities are very high allowing for detailed studies of a wide range of observables. On the other hand, lattice QCD suggests that at the (T,$\mu_B$) values probed at top RHIC energy, the transition from partonic to hadronic matter is not a phase transition but a cross over. To explore the phase transition region and the critical endpoint of QCD one clearly has to go downwards in energy. Unfortunately, decreasing the beam energy poses a problem for collider based experiments because the luminosity decreases too. Which in turn means that rare probes get out of reach. Nevertheless, a decrease in energy offers many new exciting possibilities and challenges. This is why more suitable titles for this talk may have been ''The high baryon density frontier'' or ''The quest for the critical end point'' or ''The 'where pQCD won't help you' frontier''. 

The outline of these proceedings is as follows. I will shortly review the past achievements, basically providing experimental facts of irregularities in the low energy regime between $\sqrt{s_{NN}}=5-15$~GeV. Some of these experimental result are already known, while the majority has been presented for the first time at this meeting. They all share that there is a lack of consistent interpretation of the data on the theory side. Then I discuss the next generation of experiments and facilities that may allow us to gain high precision data to explore the onset of deconfinement and the phase transition with unprecedented accuracy. Finally, I want to point out what should be done on the theory side to provide highly precise calculations for the interpretation of the experimental data. In line with my presentation, these proceedings will neglect all critical discussions. I also want to apologize to those colleagues whose exciting results could not be mentioned due to space limitations.

\section{Phase diagram and existence of the CEP}

\begin{figure} \center
\includegraphics[width=.32\textwidth]{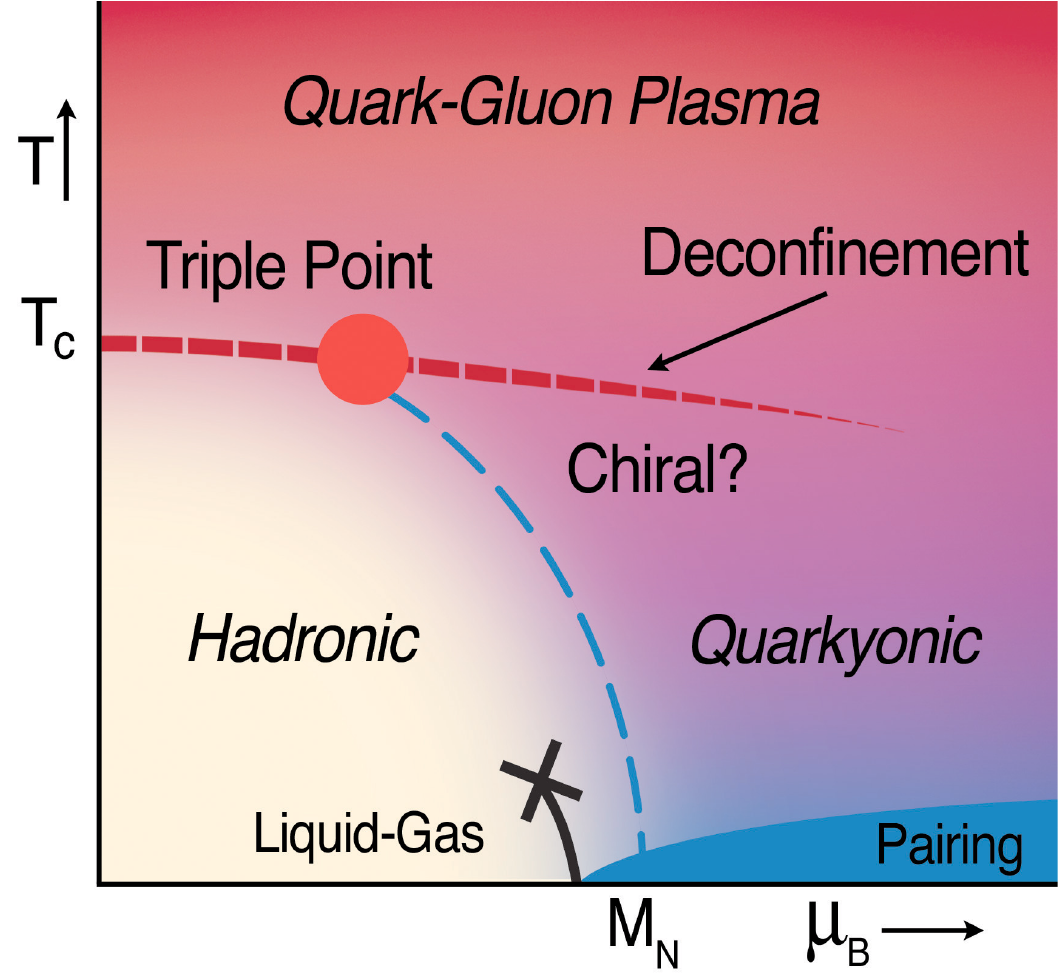}
\includegraphics[width=.32\textwidth]{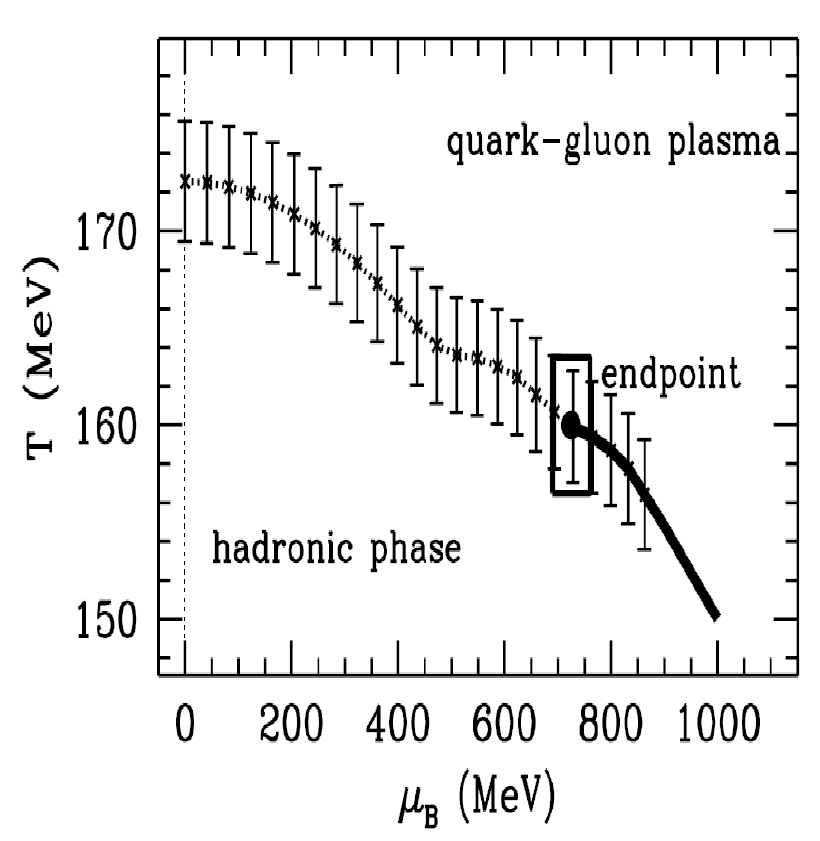}
\includegraphics[width=.32\textwidth]{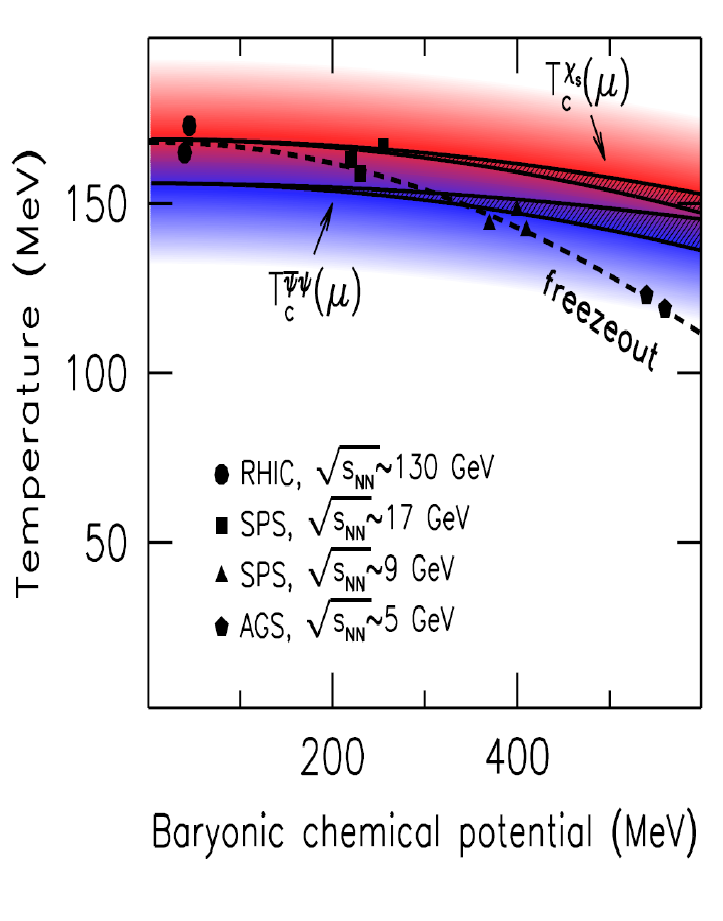}
\caption{Left: Schematic phase diagram of QCD, including the CEP, the parton-hadron transition line and the Quarkyonic phase. Center: Previous prediction for the location of the CEP from lattice QCD in the (T,$\mu_B$)-plane \cite{hep-lat/0111064}. Right: Recent prediction of the transition lines in the (T,$\mu_B$)-plane \cite{arxiv:1102.1356}.\label{fig:phasediagram}} \end{figure}

The phase diagram of QCD is depicted in Figure \ref{fig:phasediagram}. The left part of Fig. \ref{fig:phasediagram}  provides a schematic view on the qualitative features expected of QCD. Most notable are the critical endpoint (CEP) and the Quarkyonic phase. The location and existence of the critical point has been under discussion for many years \cite{STEPHANOV} (Fig. \ref{fig:phasediagram}, center, shows an early calculation in lattice QCD \cite{hep-lat/0111064}). While different lattice QCD groups have predicted the existence of a CEP \cite{hep-lat/0111064,KARSCH}, other groups have suggested that the critical surface might be bending away from the physical point \cite{arxiv:1009.4089}. Recent studies by Endroede et al, do also not provide evidence for the existence of the critical point in the investigated (T,$\mu_B$) regime below a $\mu_B$ of 600~MeV \cite{arxiv:1102.1356}.
In view of these ambiguous theoretical results, I will avoid a deeper discussion of the theoretical expectations and turn to the experimental observations in the beam energy region where a change from the first order phase transition to the cross over (via a CEP) may take place.

\section{Kinks everywhere!}
\begin{figure} \center
\includegraphics[width=.32\textwidth]{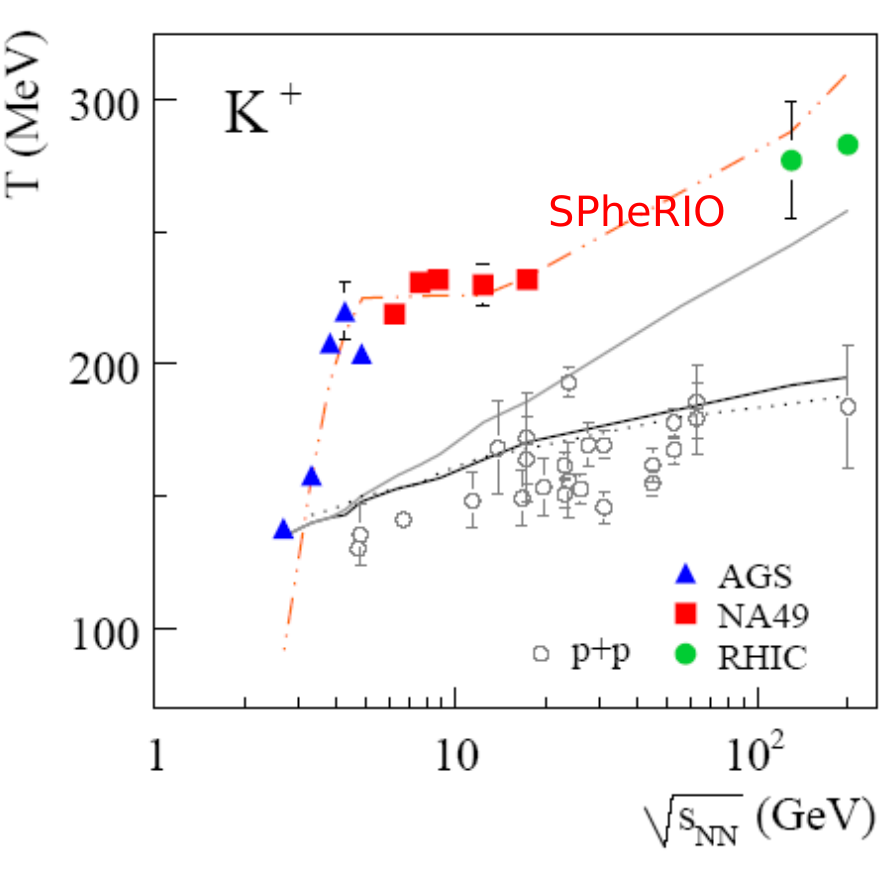}
\includegraphics[width=.32\textwidth]{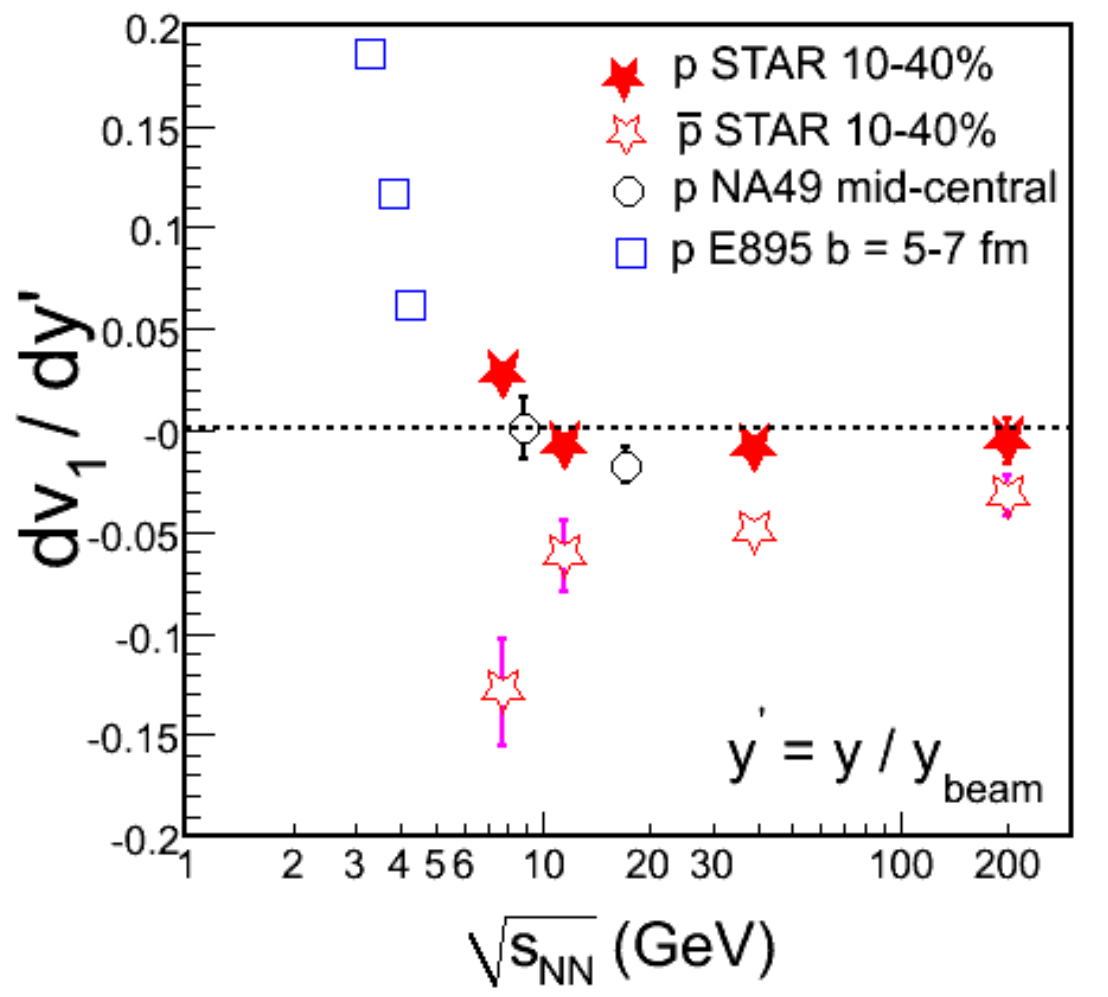}
\includegraphics[width=.32\textwidth]{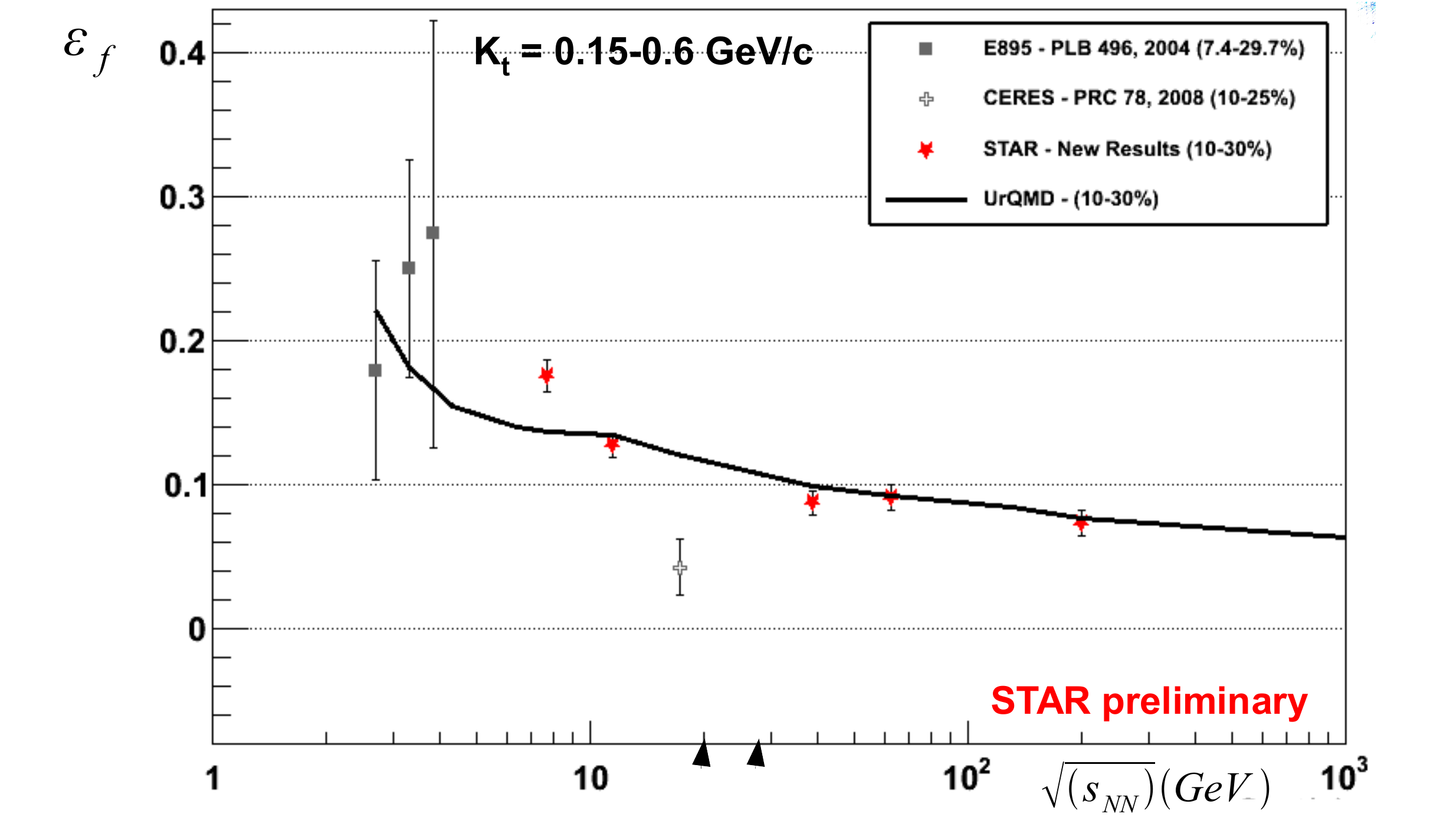}
\caption{Left: Inverse slope excitation function for Kaons \cite{SPHERIO}. Center: Flow systematics \cite{MOHANTY}. Right: Final state coordinate space eccentricity \cite{LISA}. \label{fig:kinetics}} \end{figure}

First hints for irregular and non-monotonous behaviors in energy excitation functions of nucleus-nucleus reactions were reported by the NA49 collaboration \cite{MAREK}. Among other observations is the step like structure in the inverse slope (and mean $m_T-m_0$) systematics of various particle species, see Fig. \ref{fig:kinetics} (left). This observation could be interpreted as a sign of the mixed phase which softens the equation of state (EoS). Further hints for a softening of the EoS come from the systematic study of the $v_1$ excitation function (Fig. \ref{fig:kinetics} (center)), which shows the slope of the bounce-off of protons as a function of energy. It was predicted that a sign change in the slope parameter, as now observed in the STAR data \cite{MOHANTY}, could be interpreted as a change from hadronic to partonic matter \cite{CSERNAI}. Finally let us investigate how the initial anisotropies are transformed into final state anisotropies. To this aim Fig. \ref{fig:kinetics} (right) shows the extracted final state eccentricities in coordinate space as obtained from HBT studies \cite{LISA}. Also here a non-monotonous behavior becomes visible (note the CERES point around 20 AGeV) that coincides with the energy range of the previously discussed irregular structures.   

\begin{figure} \center
\includegraphics[width=.32\textwidth]{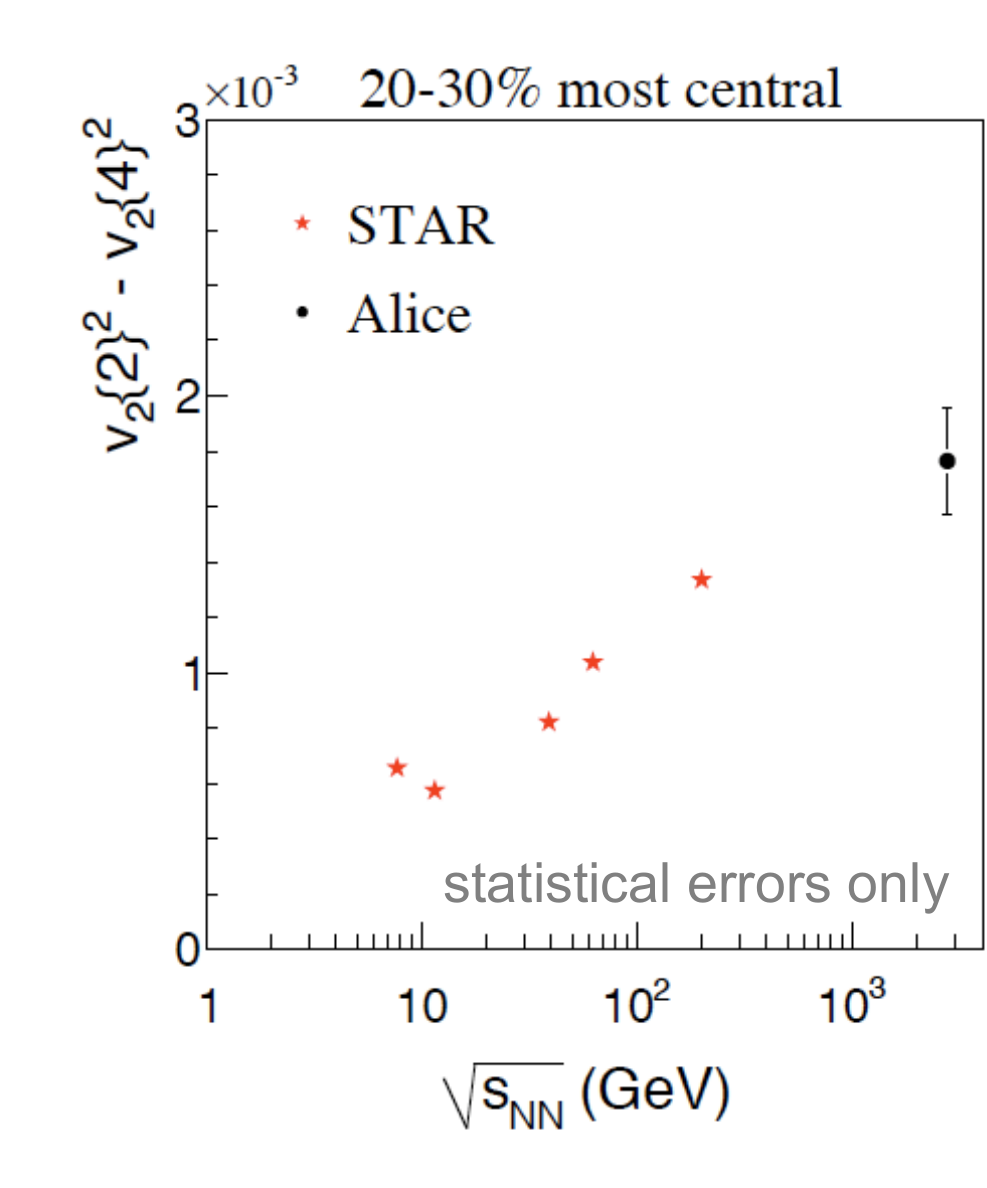}
\includegraphics[width=.32\textwidth]{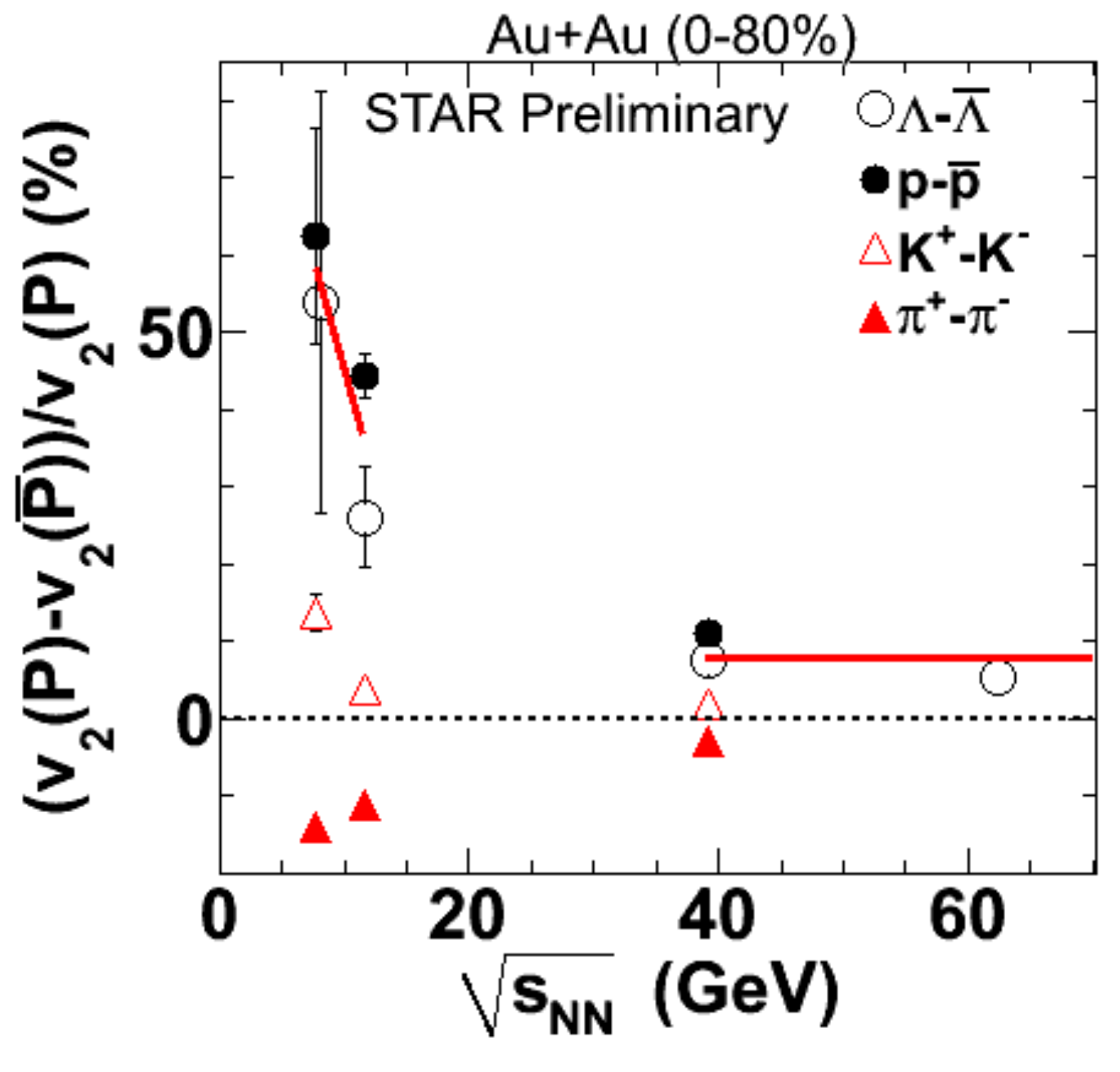}
\includegraphics[width=.32\textwidth]{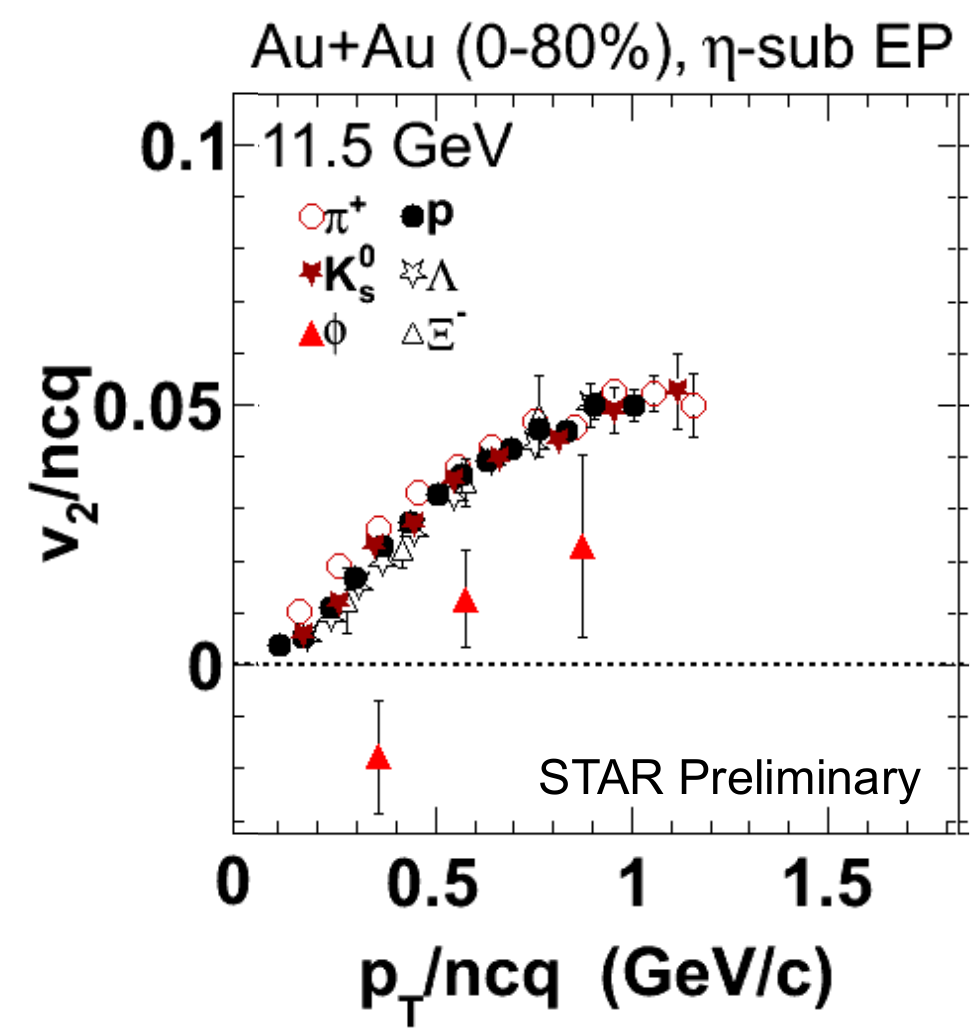}
\caption{Left: Excitation function of the non-flow contribution \cite{SORENSEN}. Center: Relative difference between the elliptic flow of particles and anti-particles \cite{SCHMAH}. Right: Scaled elliptic flow of hadrons, including the $\phi$-meson (triangles) \cite{SCHMAH}\label{fig:v2}} \end{figure}

Next let me turn to the elliptic flow analysis. After the exploratory studies at the CERN-SPS, the STAR collaborations beam energy scan (BES) program has put the breadth and the quality of these measurements to a new level. A most striking result is the excitation function of the non-flow contributions as shown in Fig. \ref{fig:v2} (left). Also here one observes a local minimum which may indicate a sudden change in the conversion efficiency of the the initial anisotropy in space into momentum space. Such a behavior may be expected if the initial state viscosity has a local minimum, e.g. due to a phase transition to a QGP. The relative importance of the hadronic stage may be inferred from the differences of the elliptic flow values of particles and anti-particles as shown in Fig. \ref{fig:v2} (center). The strong rise towards low energies can be interpreted as the emergence of a long lived hadronic state that 'eats-up' the anti-particles. Further evidence for a change in the degrees of freedom around 10~AGeV center-of-mass energy comes from the investigation of the elliptic flow of multi-strange particles. This is exemplified by the deviation of the $v_2$-values of the $\phi$-meson from the constituent quark scaling (ncq) curve (see Fig. \ref{fig:v2} (right)). If in addition, a hierarchy of the violation of the ncq-scaling could be established when going from protons to Lambdas, Xis and Omegas it could provide direct evidence for the relative importance of the hadronic phase as compared to the partonic phase in the early stage of the reaction.  

\section{Fluctuations: From lattice QCD to data}
\begin{figure} \center
\includegraphics[width=.32\textwidth]{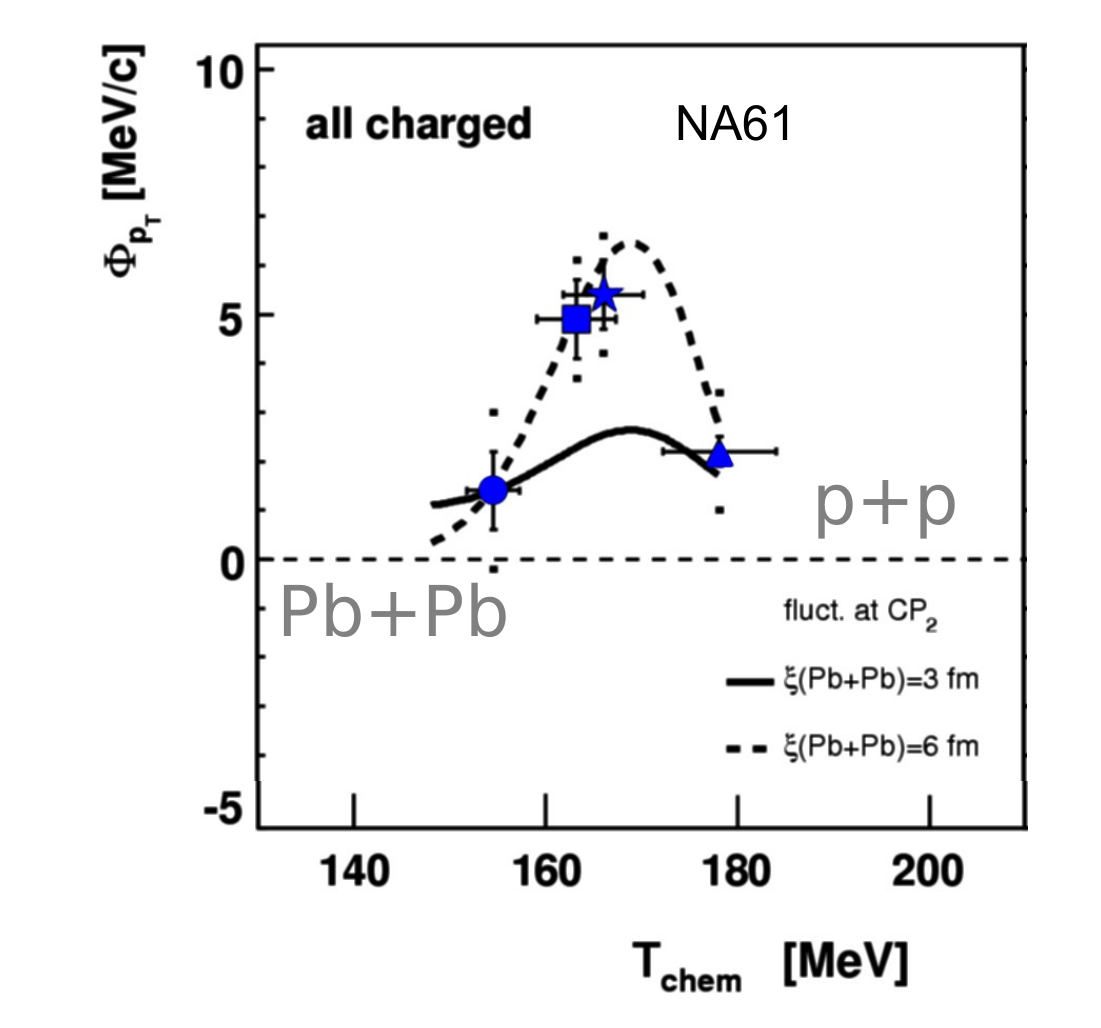}
\includegraphics[width=.32\textwidth]{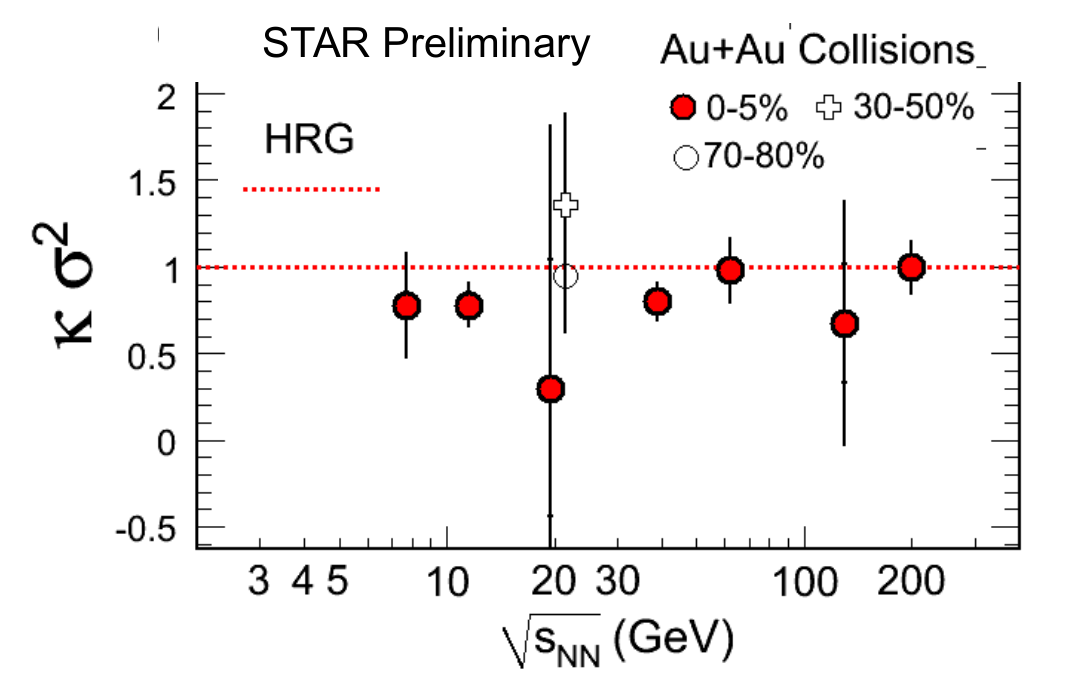}
\includegraphics[width=.32\textwidth]{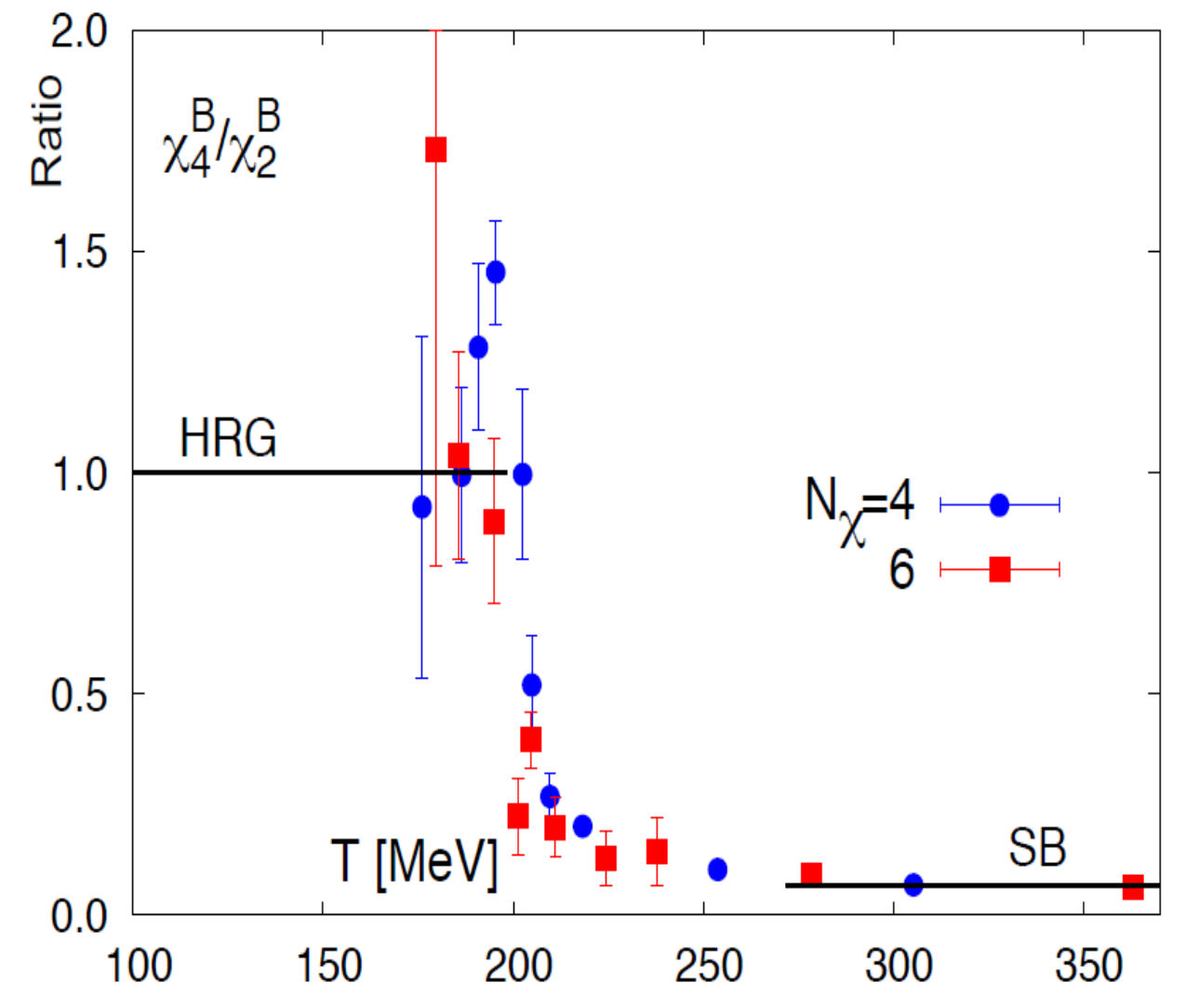}
\caption{Left: Momentum fluctuations as measured by NA61 in comparison to the expectations from an effective model at the CEP \cite{MAREK, STEPHANOV}. Center: STAR data on the scaled proton number kurtosis (as a proxy for $\chi^B_4/\chi^B_2$) \cite{MOHANTY}. Right: Lattice QCD prediction for the $\chi^B_4/\chi^B_2$-ratio \cite{CHENG}.\label{fig:flucs}} \end{figure}

While irregular structures appear in a multitude of measured data around center-of-mass energies of 5-15 AGeV, it is usually difficult to find a consistent and unambiguous theoretical interpretation. Fluctuation observables which are usually connected to well defined susceptibilities and correlation lengths might allow to gain additional insights. Very interesting data on fluctuations has been provided by the NA61 experiment (Fig. \ref{fig:flucs}, left) where the measured fluctuations are interpreted in terms of the correlation length at the critical point \cite{STEPHANOV}. If this interpretation could be confirmed by a full dynamical simulation \cite{NAHRGANG} it may provide the first experimental hint for the location of the CEP. A direct connection between data and a first principle calculation has been suggested by recent lattice QCD data \cite{CHENG} on fluctuations (Fig. \ref{fig:flucs}, right). The ratio of the 4th order to the second order baryon number susceptibilities is related to the scaled kurtosis of the event-by-event baryon number fluctuations and can be measured in experiment. These measurements have been performed by the STAR experiment and confirm the expectations, see Fig. \ref{fig:flucs}, center. Deviations from the hadron gas behavior are predicted for the ratios of even higher order susceptibilities, which modify the fluctuations below the phase transition temperature (i.e. they are in principle observable in the hadronic fluctuations) \cite{REDLICH}.

\section{Physics challenges and future perspectives}

After the previous discussions, the physics challenges in this energy regime are clear: (I) Exploration of the onset of deconfinement, i.e. the parton-hadron phase transition, the chiral phase transition and the CEP. (II) Exploration of the properties of hadrons at high baryon densities, including the Quarkyonic phase, (III) Extraction of the equation of state of QCD matter, especially the velocity of sound and the transport properties and (IV) the quest for exotica, like multi-strange objects and charmed hadrons. To this aim, high precision experimental data will be urgently needed, accompanied by high precision theoretical modeling. To meet these challenges, two new facilities will become available in the near future: FAIR (near Darmstadt, Germany) and NICA (at the JINR, Dubna, Russia). In addition, we have the currently running experimental programs at SPS (NA61/SHINE) and the RHIC-BES (STAR and PHENIX). As the impact of the RHIC-BES program and NA61 have already been discussed above, I will focus here on the potential of the new facilities for heavy ion beams.

The heavy ion program at NICA \cite{SORIN} with the MPD detector will provide collisions with light and heavy ions up to Au+Au at center-of-mass energies between $\sqrt {s_{NN}}=4-11$~GeV and luminosities around $10^{27}/{\rm cm}^2/{\rm s}$. It is supplemented by the low energy program at the Nuclotron with beam energies up to 4.5~AGeV. The physics program will focus on the onset of deconfinement with discovery potential for the CEP and the Quarkyonic phase. While the Nuclotron is already in operation, the physics start for the NICA collider program is envisaged for the year 2016.

The FAIR facility with the SIS-100 and SIS-300 program will allow to study heavy ion beams with up to 35~AGeV beam energy with ultra-high luminosities and extreme precision. The physics program will focus on the onset of deconfinement, the search for critical fluctuations, hadron properties in dense baryonic matter and exotica with an unprecedented breadth of observables. Most notably are the unique capabilities of the CBM-experiment for the study of rare probes, like charmed hadrons, multi-strange hadrons and clusters, coupled with a state-of-the-art dilepton detector. This will open up the route towards the exploration of the hadron properties near the chirally restored phase via low mass dileptons, allow for flow measurements of D-mesons and $J/\Psi$s with high statistics, discovery of MEMOs and exotic quark-gluon states and studies of (pre-cursors) of the CFL state and Quarkyonic matter via dileptons and event-by-event fluctuations. Start of the ground breaking is planned for the beginning of 2012 and the first beam is expected in the year 2017/2018 \cite{RICHTER}.

\section{Need for a joint theory effort}
Let me finally address my wish list for the developments on the theory side. Currently the theoretical approaches fall into three categories: theories in equilibrium, e.g. studies using effective Lagrangians (PNJL, quark-meson-models,\dots) but also lattice QCD, (viscous) hydrodynamic studies based on various equations of state and transport simulations (including models based on the geometrical Glauber picture). It is evident that a unified picture of the dynamical evolution of the QCD-matter will have to include features of all these approaches to allow for a consistent interpretation of all facets of the experimental data. With the extraction of transport coefficients and the EoS from lattice QCD data and their use in hydrodynamics models, we have seen a first step towards this unification \cite{HUOVINEN-PETRECKY}. On the other hand, previous studies have also shown, that one can unite hydrodynamics with transport simulations employing hybrid approaches to avoid initial stage and final stage ambiguities \cite{HYBRID}. However, it is possible to go even further and simulate the hydrodynamical behavior using only a Boltzmann approach \cite{CARSTEN}. An orthogonal approach with yet not fully recognized potential are real-time (as compared to imaginary time) lattice QCD studies \cite{NARA-DUMITRU}. Here one has the potential to obtain a real dynamical evolution of the QCD system based on first principles. 

From my point of view, the ultimate goal of these developments should be to design a single and open standard model for the description of (heavy) ion reactions, including a deconfinement transition with the proper order, including the dynamics near the CEP, chiral symmetry with its breaking and restauration,  multi-particle interactions and off-shell dynamics. This goal can only be reached as a joined community effort. It will be worthwhile to pursue because it will provide a solid basis for the interpretation of the experimental data. First attempts for joined theory activities have already started with the MADAI collaboration and the TechQM initiative and should be broadened and internationalized.

\section{Summary and outlook}
I have discussed the status of the low energy regime for heavy ion reactions where the location of the critical end point and the onset of deconfinement is expected. A large body of experimental data has become available during this conference which has confirmed and tremendously extended the previously observed irregularities and highly interesting results from the CERN-SPS. Most noteworthy to me are the elliptic flow studies for $\phi$-mesons and the $v_1$, and $v^2_2(2)-v_2^2(4)$ studies at the RHIC-BES which provide many additional insights and even more questions. Currently, the field is in a very comfortable situation with dedicated running programs (RHIC-BES and NA61) that provide pioneering studies geared to pin down the CEP and the onset of deconfinement. The future is bright with two upcoming facilities that will allow to push these investigations to a new level in terms of data quantity and quality. These facilities will allow to explore the properties of QCD matter in the respective (T,$\mu_B)$ regions with novel and dedicated probes like charm and dileptons which will provide challenges and results for the next two decades. The ultimate goal will be to obtain a concise and unambiguous interpretation of the experimental data in terms of the equation of state and the transport properties of the QCD matter at high baryon densities. This will need a coordinated and joined theory activity to model and understand the experimental results not only in the large wave length limit (i.e. hydrodynamics), but also to understand what actually makes up this liquid on the microscopic level.

\section*{Acknowledgements} This work was supported by the Helmholtz
International Center for FAIR within the framework of the LOEWE program launched
by the State of Hesse, GSI, and BMBF.

\section*{References}

\end{document}